\documentclass{b98proc_m}

\def\lsim{\mathrel{\rlap{\lower4pt\hbox{\hskip1pt$\sim$}}
    \raise1pt\hbox{$<$}}}         
\def\gsim{\mathrel{\rlap{\lower4pt\hbox{\hskip1pt$\sim$}}
    \raise1pt\hbox{$>$}}}         
\def\Journal#1#2#3#4{{#1} {\bf #2}, #3 (#4)}

\def\PLB{{\em Phys. Lett.} B}

\def\PRL{\em Phys. Rev. Lett.}

\def\PRD{{\em Phys. Rev.} D}

\def\AJ{\em Astrophys. J.}
\def\PRP{\em Phys. Rep.}

\begin{document}

\title{
THEORETICAL ISSUES IN NEUTRINO PHYSICS
}

\author{W. C. HAXTON}
\address{Institute for Nuclear Theory, Box 351550, and Department of
Physics, Box 351560, \\ 
University of Washington, Seattle, WA 98195, USA}


\maketitle

\abstracts{
I review a number of the open questions about neutrino properties,
critique recent hints of neutrino mass, and discuss one recently
proposed neutrino mass matrix to illustrate the direction in
which we may be headed.  I also present one example of the 
implications of these new developments for astrophysics.
}

\section{Introduction}

\noindent
In this talk I would like to discuss a number of the open questions
we have entertained about neutrinos since they were first postulated
more than 65 years ago, as well as a few of the possible answers
that may result from atmospheric, solar, and terrestrial neutrino
experiments.  Indeed, the list of open questions has proven
surprisingly resistant to experiment:\\
$\bullet$ What are the masses of the $\nu_e$, $\nu_\mu$, and
$\nu_\tau$, and what will these masses tell us about new scales
beyond the standard model? \\
$\bullet$ Are there massive neutrinos beyond the range tested by
measurements of the $Z^0$ width? \\
$\bullet$ What are the particle-antiparticle conjugation properties
of the known neutrinos? \\
$\bullet$ Do neutrinos have nonzero electromagnetic moments
(magnetic dipole, electric dipole, or anapole), or perhaps a nonzero
charge radius? \\
$\bullet$ Do neutrinos of different flavor mix to produce neutrino
oscillations?\\
$\bullet$ Can we prove cosmic background neutrinos exist?  What
role have they played in determining the present structure of our
universe? 
Do they comprise an appreciable fraction of the dark matter? \\
$\bullet$ What is the role of neutrinos in core-collapse supernovae?
Are nonstandard neutrino properties essential to the explosion mechanism or to the associated
nucleosynthesis (r-process, $\nu$-process)? \\
$\bullet$ What are the sources of very high energy neutrinos in
astrophysics?  Are they associated
with gamma ray bursts or active galactic nuclei?  Does the
standard model properly describe their propagation through and interactions
with the cosmic background radiation? \\
  
\noindent
The above list could be extended for several more pages.  What we
do know about neutrinos is, in some respects, equally puzzling.
For example direct mass limits tell us that 
\begin{center}
m($\nu_e) \lsim (3-5)$ eV \cite{emass} \\
m($\nu_\mu) \lsim 170$ keV \cite{mumass} \\
m($\nu_\tau) \lsim 18.2$ MeV \cite{mumass} \\
\end{center}
Naively, such small values pose a problem for theorists hoping
to extend the standard model by unifying the known particles 
into larger multiplets.  For example, an attractive idea
\cite{su5} might be multiplets containing all of the particles
of a given family
\[ ~\left( \begin{array}{c} \nu \\ e \end{array} \right) 
\begin{array}{c} e.g. \\ \longrightarrow \\ grander \\ model
\end{array} \left( \begin{array}{c} u \\ d \\ \nu \\ e \end{array}
\right) \]
One then might expect (to within group theory factors)
that the members of the multiplet would couple to the mass-generating
fields in a similar way, and thus have about the same mass,
$m_i \sim o(m_D)$.
Now the u and d quarks and the electron have masses on the order
of an MeV, but the $\nu_e$ clearly breaks the pattern: it is at 
least six orders of magnitude lighter.  One popular resolution of
this dilemma is connected with additive quantum numbers.  For
example 
\[ e^- \rightarrow e^+ \] 
under charge conjugation, clearly producing an orthogonal
antiparticle, distinguished from the electron by an additive
quantum number (the charge).  However the corresponding 
question for the $\nu$ --- does there exist a distinct 
antiparticle, $\bar{\nu}$? --- is not so easy to answer. \\

\noindent
Before tackling this question, it is helpful to discuss some of
the complicating issues connected with the handedness of 
massive neutrinos.  Consider a massive left-handed neutrino
moving at a velocity $v < c$.  Now boost the observer
to frame moving faster than
that neutrino's velocity
\[ \nu_{LH}  \begin{array}{ccc}
\stackrel{\leftarrow}{s} & ~~~~~\longrightarrow~~~~~ 
& \stackrel{\leftarrow}{s} \\
 \stackrel{\rightarrow}{p} & ~~~~~boost~~~~~ &
\stackrel{\leftarrow}{p} \end{array} \nu_{RH} \] 
In the boosted frame the neutrino is right-handed.
From this exercise we learn that the Lorentz structure of any 
model describing massive neutrinos demands both $\nu_{LH}$ and
$\nu_{RH}$.  As the interactions of the standard model are 
V-A, a closely associated point is that neutrino masses break
the ``$\gamma_5$" invariance of interactions, leading to 
interesting effects of order $(m_\nu / E_\nu)$: we will see an
illustration of this in a later discussion of double beta decay. \\

\noindent
It would seem that the logical way to resolve the issue of a 
$\bar{\nu}$ distinguishable from the $\nu$ is to test the 
properties of these particles experimentally.  If 
neutrinos are produced by a $\beta^+$ source and their interactions 
tested in a target downstream, one finds
\begin{center}
~~e$^+$~~~~~~~~~~~~~~~~~~~~~~~$\nu_e$~~~~~~~~~~~~~~~~~$\nu_e$~~~~~~~~~~~~~~~~~~~~~~e$^-$ \\
$\leftarrow$ \hspace{-.22in} \rule[.025in]{.5in}{.02in}
\hspace{-.09in} \fbox{\rule[-.4in]{0in}{1in} $\beta^+$ source}
\hspace{-.09in} \rule[.025in]{.5in}{.02in}
\hspace{-.22in} $\rightarrow$
~~$\cdots$~~
\rule[.025in]{.5in}{.02in} \hspace{-.22in} $\rightarrow$ 
\hspace{-.09in} \fbox{\rule[-.4in]{0in}{1in} ~target~}
\hspace{-.09in} \rule[.025in]{.5in}{.02in}
\hspace{-.22in} $\rightarrow$ \\
~~~~~~$\nu_e \equiv$ e$^+$ partner~~~~~~~~~~~~~~~~~~~~~~
always produces an e$^-$
\end{center}
That is, if we define a $\nu_e$ as the partner of the e$^+$ in
a $\beta$ decay reaction, then we observe that $\nu_e$s always
produce $e^-$s when they react in a target.  Similarly
\begin{center}
~~e$^-$~~~~~~~~~~~~~~~~~~~~~~~$\bar{\nu}_e$~~~~~~~~~~~~~~~~~$\bar{\nu}_e$~~~~~~~~~~~~~~~~~~~~~~e$^+$ \\
$\leftarrow$ \hspace{-.22in} \rule[.025in]{.5in}{.02in}
\hspace{-.09in} \fbox{\rule[-.4in]{0in}{1in} $\beta^-$ source}
\hspace{-.09in} \rule[.025in]{.5in}{.02in}
\hspace{-.22in} $\rightarrow$
~~$\cdots$~~
\rule[.025in]{.5in}{.02in} \hspace{-.22in} $\rightarrow$ 
\hspace{-.09in} \fbox{\rule[-.4in]{0in}{1in} ~target~}
\hspace{-.09in} \rule[.025in]{.5in}{.02in}
\hspace{-.22in} $\rightarrow$ \\
~~~~~~$\bar{\nu}_e \equiv$ e$^-$ partner~~~~~~~~~~~~~~~~~~~~~~
always produces an e$^+$
\end{center}
where the $\bar{\nu}_e$ has been defined as the partner of the $e^-$
in $\beta$ decay.  It would appear that the $\nu_e$ and
$\bar{\nu}_e$ so defined are operationally distinct
\begin{center}
$\nu_e$s produce e$^-$s \\
$\bar{\nu}_e$s produce e$^+$s
\end{center}
This motivates the introduction of a distinguishing 
quantum number (lepton number)
\[ \begin{array}{cc}
\underline{lepton} & \underline{l_e} \\ \mathrm{e}^- & +1 \\ \mathrm{e}^+ & -1 \\
\nu_e & +1 \\ \bar{\nu_e} & -1 \end{array} \]
If we require that lepton number is additively conserved,
\[ \sum_{in} l_e = \sum_{out} l_e, \]
the ``experimental" results discussed above then follow. \\ 

\begin{figure}[ht]
\psfig{bbllx=-6.0cm,bblly=1.5cm,bburx=18cm,bbury=21.0cm,figure=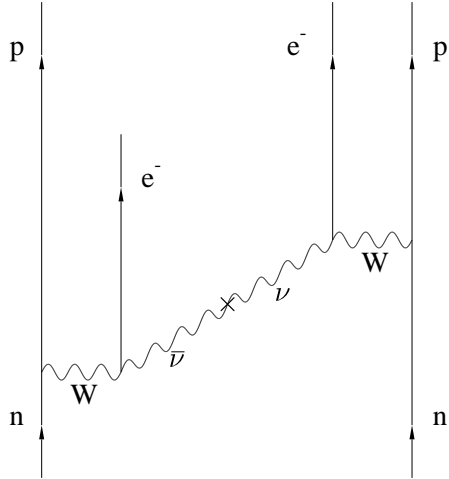,height=2.7in}
\caption{Two-nucleon diagram for neutrinoless $\beta \beta$ decay.
The amplitude vanishes if the $\nu$ and $\bar{\nu}$ are distinct,
i.e., carry different lepton numbers.}
\label{one}
\end{figure}

\noindent
The experiments described above are done by nature virtually in 
the process of neutrinoless $\beta \beta$ decay,
(A,Z) $\rightarrow$ (A,Z+2) + e$^-$ + e$^-$, as depicted in
Fig.~\ref{one}.  The illustrated process
cannot take place if the emitted
neutrino is orthogonal to the antineutrino that must be absorbed
on the second nucleon
\[ \nu_e(l = +1)~ \bot ~\bar{\nu}_e(l = -1). \] 
Results for the current generation of enriched $^{76}$Ge $\beta \beta$
decay searches \cite{ge}
\[ \tau_{1/2}(\mathrm{Ge}^{76}) \gsim 2 \cdot 10^{25} \mathrm{y} \]
imply a limit on the electron neutrino Majorana mass of $\langle \mathrm{m}_\nu^{\mathrm{Maj}}\rangle   
\lsim 0.4$ eV, where \cite{haxtonbb}
\[ \langle \mathrm{m}_\nu^{\mathrm{Maj}} \rangle =
\sum_{i=1}^{2n} \eta^{\mathrm{CP}}_i \mathrm{U}_{ei}^2 \mathrm{m}_i. \] 
Here $m_i$ is the mass of the $ith$
eigenstate, $U_{ei}$ is the amplitude of that mass eigenstate 
in $|\nu_e \rangle$, and $\eta_i^{CP}$ is the relative CP
of the $ith$ mass eigenstate.  Thus CP conservation has been assumed. \\
  
\noindent
While the nonobservation of neutrinoless $\beta \beta$
decay is consistent with a Dirac neutrino---one where the $\nu$
and $\bar{\nu}$ are distinguished by their opposite lepton 
numbers---it is not required, due to the V-A character of 
standard model weak interactions.  The replacements
\[ \nu_e \rightarrow \nu_e^{LH} \]
\[ \bar{\nu}_e \rightarrow \nu_e^{RH} \]
lead to a helicity missmatch in the above diagram and a decay
rate suppressed by o$({\langle \mathrm{m}_\nu^{Maj} \rangle \over \mathrm{E}_\nu})^2$.
Thus the absence of neutrinoless $\beta \beta$ decay is 
consistent with a Majorana neutrino ($\nu = \bar{\nu}$) if the
electron neutrino Majorana mass is light, as indicated above. \\
  
\section{The Neutrino Mass Matrix \protect\cite{haxtonbb,kayser}}
\noindent
A Majorana $\nu_e$ corresponds to the limiting case where a 
state of definite mass has two components, with both the 
boosts and particle-antiparticle conjugation (CPT, or CP in the limit of
CP conservation) coupling one component to the other, as depicted
below.  A Dirac $\nu_e$ is a four-component neutrino, where
the $\nu$ and $\bar{\nu}$ are distinguished by their lepton
numbers, and thus where the boosts and CP/CPT operations 
connect distinct components.
\begin{center}
boost\\
\rule[-.12in]{.02in}{.12in} \hspace*{-.11in} \rule{2in}{.02in}
\hspace*{-.11in} \rule[-.12in]{.02in}{.12in} \\
$\nu_{\mathrm{LH}}$~~~~~~~~~~~~~~~~~~~~~~~~~~~~~~~~~~~~$\nu_{\mathrm{RH}}$ \\
\rule{.02in}{.12in} \hspace*{-.11in} \rule{2in}{.02in}
\hspace*{-.11in} \rule{.02in}{.12in} \\
CPT\\
\end{center}
   
\begin{center}
boosts\\
\rule[-.3in]{.02in}{.3in} \hspace*{-.09in} \rule{1in}{.02in}
\hspace*{-.11in} \rule[-.3in]{.02in}{.15in}
\hspace*{-.11in} \rule[-.15in]{2in}{.02in}
\hspace*{-.11in} \rule[-.3in]{.02in}{.15in} \hspace*{-2.09in}
\rule{3in}{.02in} \hspace*{-.11in} \rule[-.3in]{.02in}{.3in} \\
$\nu_{\mathrm{LH}}$~~~~~~~~~~~~~~~~~~$\bar{\nu}_{\mathrm{RH}}$~~~~~~~~~~~~
~~~~~~~~~~~~~~~~~~~~~~$\bar{\nu}_{\mathrm{LH}}$~~~~~~~~~~~~~~~~~~$\nu_{\mathrm{RH}}$ \\
\rule{.02in}{.15in} \hspace*{-.11in} \rule{1in}{.02in}
\hspace*{-.11in} \rule{.02in}{.15in}
\hspace*{1.89in} \rule{.02in}{.15in} \hspace*{-.11in} 
\rule{1in}{.02in} \hspace{-.11in} \rule{.02in}{.15in} \\
CPT~~~~~~~~~~~~~~~~~~~~~~~~~~~~~~~~~~~~~~~~~~~~~~~~~~~~~~~~CPT
\end{center}

\noindent
We now proceed through a simple exercise of generalizing these
limiting cases to one where several mass eigenstates may 
contribute, and where both Dirac and Majorana mass terms
arise.  The starting point is a Dirac field from which we 
project the four components, using the R/L
and charge conjugation projection operators:
\begin{center}
$\psi_{R/L} = {1 \over 2} (1 \pm \gamma_5) \psi$ \\
C$\psi_{R/L}$C$^{-1} = \psi^{c}_{R/L}$ 
\end{center}
We apply this to the mass term for the Dirac equation
\[ \mathcal{L}_m(x) \sim m_D \bar{\psi}(x)\psi(x)  
 \Rightarrow M_D 
\bar{\Psi}(x)\Psi(x) \]
where $m_D$ has been replaced by a nondiagonal 
$3 \times 3$ matrix $M_D$ in flavor space and
\[ \Psi = \left( \begin{array}{c} \psi^e \\ \psi^\mu \\
\psi^\tau \end{array} \right) \]
The resulting mass matrix
\[ \begin{array}{c} (\bar{\Psi}^c_L,\bar{\Psi}_R,
\bar{\Psi}_L,\bar{\Psi}^c_R) \\ \\ \\ \end{array}
\left( \begin{array}{cccc} 0 & 0 & 0 & M^T_D \\
0 & 0 & M_D & 0 \\ 0 & M_D^\dag & 0 & 0 \\ M_D^* & 0 & 0 & 0
\end{array} \right) 
\left( \begin{array}{c} \Psi^c_L \\ \Psi_R \\ \Psi_L \\ \Psi^c_R
\end{array} \right) \]
then allows for flavor oscillations, as 
$M_D$ is assumed to be nondiagonal. \\

\noindent
While the upper left and lower right quadrants of this matrix
must be zero because the left- and right-handed projectors
annihilate each other, obviously additional terms can be
introduced elsewhere if we respect the requirement of
hermiticity.  Specifically,
\[ \mathcal{L}_m(x) \Rightarrow M_D
\bar{\Psi}(x)\Psi(x)
+\bar{\Psi}^c_LM_L\Psi_L + \bar{\Psi}^c_RM_R\Psi_R \]
so that the mass matrix becomes
\[ \begin{array}{c} (\bar{\Psi}^c_L,\bar{\Psi}^R,
\bar{\Psi}_L,\bar{\Psi}^c_R) \\ \\ \\ \end{array}
\left( \begin{array}{cccc} 0 & 0 &  M_L & M^T_D \\
0 & 0 & M_D &  M_R^\dag \\  M_L^\dag & M_D^\dag & 0 & 0 \\ M_D^* &  M_R & 0 & 0
\end{array} \right) 
\left( \begin{array}{c} \Psi^c_L \\ \Psi_R \\ \Psi_L \\ \Psi^c_R
\end{array} \right) \]
The new Majorana mass terms break the local gauge invariance $\psi(x) \rightarrow
e^{i\alpha(x)} \psi(x)$ associated with a conserved lepton number.
It is these nonDirac mass terms that can generate the nonzero
$\langle m_\nu^{Maj} \rangle$ that gives rise to neutrinoless
$\beta \beta$ decay. \\

\noindent
One can proceed to diagonalize this matrix
\[ \Psi_{\nu(e)}^L = \sum_{i=1}^{2n} U_{ei}^L \tilde{\nu}_i(x) 
~~\mathrm{with~masses}~m_i \]
The eigenstates are two-component Majorana neutrinos \cite{haxtonbb},
yielding the proper $2 \times 2n = 4n$ degrees of freedom,
where $n$ is the number of flavors.
We can recover the Majorana and Dirac limits:\\
$\bullet$ If $M_R$ = $M_L$ = 0, the eigenstates of this matrix
become pairwise degenerate, allowing the $2n$ two-component 
eigenstates to be paired to form $n$ four-component Dirac
eigenstates.\\
$\bullet$ If $M_D$ = 0, the left- and right-handed components
decouple, yielding $n$ left-handed Majorana eigenstates with
standard model interactions. 
  
\noindent
There are interesting physical effects associated with these 
limits.  Dirac neutrinos can have magnetic dipole, electric
dipole (CP and T violating), and anapole (P violating) moments,
as well as nonzero charge radii.  Majorana neutrinos can have
anapole moments but only transition magnetic and electric
dipole moments.  Yet transition moments are quite interesting
in the context of matter-enhanced spin-flavor oscillations
\cite{akhmedov}.
The most stringent limits on both diagonal and transition magnetic
and electric dipole moments come from red giant evolution, 
where the enhanced neutrino pair production delays core He
ignition.  This yields \cite{raffelt}
\[ |\mu_{ij}| \lsim \mathrm{few} \cdot 10^{-12} \mu_B \]
a bound that is approximately two orders of magnitude more restrictive
than the laboratory limit. \\
  
\noindent
Neutrinos are unique among the fermions in allowing both Dirac
and Majorana mass terms, a consequence of the absence of any 
obvious additive quantum numbers that must change under
particle-antiparticle conjugation.  The presence of both mass
terms provides an attractive explanation for small neutrino masses,
the seesaw mechanism of Gell-Mann, Ramond, Slansky, and
Yanagida \cite{seesaw}.  As $\beta \beta$ decay suggests a
left-handed Majorana mass much smaller than typical Dirac masses,
while right-handed interactions are not seen at low energies
and thus might be characterized by mass scales well beyond 
the standard model, the following mass matrix is natural:
\[ \left( \begin{array}{cc} 0 & m_D \\ m_D & m_R \end{array} \right)
\Rightarrow m^{light}_\nu = m_D ( {m_D \over m_R} ) \]
Thus $m_D/m_R$ is the needed small parameter explaining why
neutrinos are so much lighter than their charged partners.
If the $\nu_\tau$ mass is on the order of 0.1 eV (a value suggested
by atmospheric neutrinos), and the Dirac mass is taken from
$m_{top} \sim 180$ GeV, this yields $m_R \sim 0.3 \times 10^{15}$
GeV, a value reasonably close to the GUT scale, $M_{GUT} \sim
10^{16}$.  The massive
right-handed neutrino of the seesaw fits naturally into various
extended models \cite{su5,wilczek}.  For example, the
16-dimensional family multiplet of SO(10) is
\[ (\vec{\mathrm{u}}_L,\vec{\mathrm{d}}_L,
\vec{\mathrm{u}}_R,\mathrm{e}_R,\vec{\mathrm{d}}_R,\mathrm{e}_L,
\nu_L,\nu_R). \]
That is, the assignment provides a natural spot to be filled by
a heavy, chargeless, right-handed neutrino.  Thus the key question
is whether current hints of neutrino mass are telling us about
physics at $10^{15}$ GeV. \\
   
\section{Handicapping the Hints of Mass \protect\cite{robertson}}
\begin{figure}[ht]
\psfig{bbllx=-0.7cm,bblly=4.5cm,bburx=12cm,bbury=24.3cm,figure=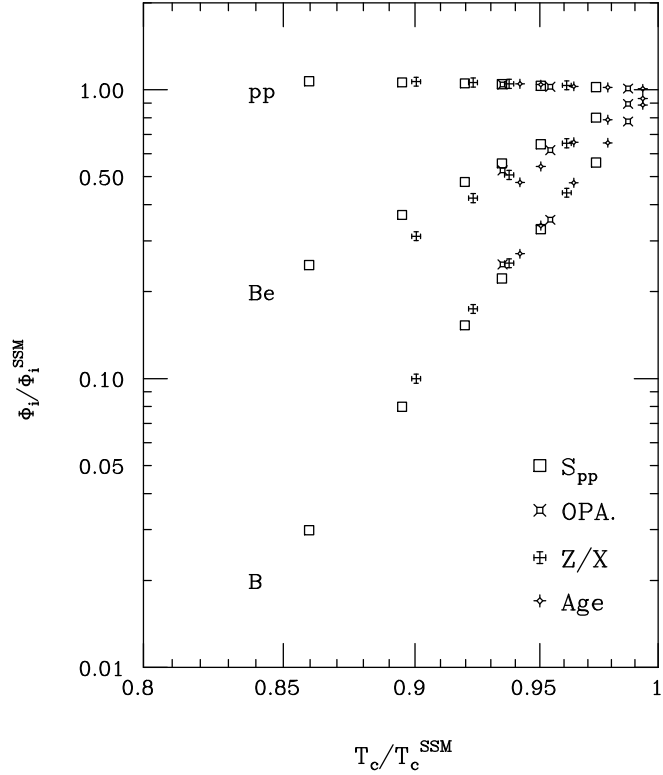,height=4in}
\caption{The response of the pp, $^7$Be, and $^8$B solar neutrino
fluxes to the indicated variations in solar model input 
parameters, displayed as a function of the resulting central
temperature $T_c$.  From \protect\cite{castellani}.}
\label{two}
\end{figure}

\noindent
The solar neutrino problem, the discrepancy between the predictions
of the standard solar model (SSM) \cite{bahcall} and the results of
the $^{37}$Cl, GALLEX and SAGE, and KamiokaII/III and
SuperKamiokande experiments, was described by Hamish Robertson.
There are two somewhat distinct arguments that this discrepancy
requires new particle physics: \\
$\bullet$ There is the $\sim 3 \sigma$ argument based on 
global fits to the various experiments, using undistorted 
neutrino spectra but making no other assumptions, or only 
weak assumptions (such as a steady state model where the 
luminosity constrains present fusion rates), about the solar
model (see, for example, \cite{heeger}).  In such fits, an
unphysical result is obtained, a negative $^7$Be flux.\\
$\bullet$ There is also a $\sim 5 \sigma$ argument in which
the general temperature dependence of standard and nonstandard
models is used.  This requires some additional assumptions,
such as solar burning under the conditions of chemical
equilibrium \cite{cumming}, but is still independent of
many details of the solar model.  The conclusion from 
experiment that $\phi(^8$B) $\sim 0.4 \phi^{SSM}(^8$B) and
from solar models that $\phi(^8$B) $\propto T_c^{18}$, where
$T_c$ is the core temperature, leads to the conclusion
\[  T_c \sim 0.96 T_c^{\mathrm{SSM}} \]
That is, a cool solar core is required. Consequently as
\[ \Phi(^7\mathrm{Be})/\Phi(^8\mathrm{B}) \sim
T_c^{-10} \]
such a cooler core requires
\[ {\Phi(^7 \mathrm{Be}) \over \Phi(^8 \mathrm{B})} \sim
1.5 {\Phi^{\mathrm{SSM}}(^7 \mathrm{Be}) \over \Phi^{\mathrm{SSM}}(^8 \mathrm{B})} \]
But experimentally we find $\Phi(^7$Be) $\sim$ 0,
so that the experimental ratio is low, as would be expected for a hotter
core.  This is the crux of the second argument: experiment 
leads to a contradiction in the absence of new particle physics,
with one observation ($\Phi(^8$B)) requiring a cooler core and a second ($\Phi(^7$Be)/$\Phi(^8$B)) a
hotter one.  This conflict is nicely illustrated in Fig.~\ref{two},
taken from Castellani et al. \cite{castellani}. \\
  
\noindent
These two indirect arguments are quite convincing,
leading many in the field to conclude that the solar neutrino
problem demands new particle physics.
Yet, as the conclusions are based on combining the results from several
experiments, no one of which by itself requires new physics,
it would be very reassuring to see a ``smoking gun" signal.
The current generation of active detectors -
SuperKamiokande with its spectral sensitivity and SNO with its
direct sensitivity to neutral currents - has the potential to
yield such a signal.  Currently the SuperK spectral
distribution does show some structure, but of an unexpected
kind: there is a surprising excess of high-energy electrons,
similar to those expected from the $^3$He + p reaction if the
standard cross section estimate were to be multiplied by about a factor of 30 \cite{nu98}. \\

\noindent
Experiments sensitive to atmospheric neutrinos have traditionally
expressed their results in terms of a ratio of ratios
\[ R = {(N_\mu/N_e)_{DATA} \over (N_\mu/N_e)_{MC}} \]
determined from the measured and calculated (with Monte Carlo codes)
muon-like and
electron-like neutrino rates.  Robertson's talk gave many of the
reasons the atmospheric neutrino problem is considered such
a convincing argument for new physics: \\
$\bullet$The SuperK ratio has been very accurately determined, $\sim 0.61 \pm 0.03 \pm 0.05.$ \\
$\bullet$There is good consistency between sub-GeV/multi-GeV 
and fully contained/partially contained data sets. \\
$\bullet$ The change in the ratio with zenith angle provides
direct evidence for neutrino oscillations. \\
$\bullet$ The results for $R$ are consistent among the four 
detectors with the largest data sets (SuperK, SoudanII, IMB,
Kamiokande). \\
$\bullet$ The up-down difference is ``self-normalizing," almost
independent of the calculated atmospheric fluxes. \\
  
\noindent
The favored interpretation of the SuperK and other atmospheric
neutrino results is a large-mixing-angle $\nu_\mu \rightarrow \nu_\tau$
oscillation.  There have been some questions raised about this
interpretation: \\
$\bullet$ The absolute rates can be fit as well by an excess of 
e-like events as by a deficit in $\mu$-like events.
\[ \begin{array}{ccc} & data & Monte Carlo \\ 
e-like & 983 & 812 \\ \mu-like & 900 & 1218 \end{array} \]
$\bullet$The SuperK sin$^22\theta - \delta m^2$ region favors smaller 
values of $\delta m^2$ than those found by Kamioka and SoudanII, 
though there is a region of overlap. \\
$\bullet$ There is some tension between the SuperK shape fit and $R$,
with the results for $R$ largely agreeing with other atmospheric
neutrino experiments, but lying mostly outside the region favored
by the shape fit. \\
$\bullet$ The results require very large mixing angles, with more
than half of the 90\% confidence level region lying in the
unphysical region ($\sin^2 2 \theta$ exceeding 1) in an unconstrained fit.\\
Yet despite these concerns, the most
striking aspect of the SuperK results is the azimuthal
dependence, which is direct evidence for neutrino oscillations.
And, as will be illustrated below, a scenario with maximal 
mixing angles can be made to fit nicely with other hints of
neutrino mass. \\
  
\noindent
There is a third hint of neutrino mass, direct laboratory
evidence from
the LSND experiment, although the allowed region is mostly 
excluded by the KARMEN experiment.  It is very difficult to
evaluate this situation.  If LSND were proven to be correct,
its inclusion with the solar and atmospheric neutrino problems
very much strains three-flavor fits to these three results. \\
  
\noindent
\section{Prejudices and One Possible Pattern}
\noindent
To provide some picture of how these various results might fit
together to form some pattern, I now discuss a recent 
paper by Georgi and Glashow \cite{georgi98}.  The assumptions
of their construction are:\\
$\bullet$ Three light Majorana neutrinos \\ 
$\bullet$ The atmospheric neutrino problem is due to 
$\nu_\mu \rightarrow \nu_\tau$ oscillations, since the 
$\nu_\mu \rightarrow \nu_e$ alternatively is ruled out by the
Chooz experiment. \\
$\bullet$This oscillation is nearly maximal with 
sin2$\theta_{23} \sim$ 1 and
5 $\cdot 10^{-4}$ eV$^2 \lsim \delta m_{23} \lsim 6 \cdot 10^{-3}$ eV$^2$. \\
$\bullet$The solar neutrino problem is due to oscillations with
$6 \cdot 10^{-11}$ eV$^2 \lsim \delta m^2 \lsim 2 \cdot 10^{-5}$ eV$^2$. \\
$\bullet$ The neutrino masses are constrained to satisfy
m$_1$+m$_2$+m$_3 \sim$ 6 eV in order to generate hot dark 
matter for large scale structure formation (a somewhat 
speculative condition). \\
$\bullet$The absence of neutrinoless double $\beta$ decay requires  
$\langle \mathrm{m}^{Maj}_\nu \rangle \lsim 0.4 \mathrm{eV}$,
so choose $\langle \mathrm{m}^{Maj}_\nu \rangle \sim$ 0. \\ 
$\bullet$Because of the LSND/KARMEN conflict, the LSND results 
are not considered. \\
  
\noindent
These constraints lead to a pattern of three nearly degenerate
massive neutrinos with m$_i \sim M$ and a simple mass matrix
that accounts for the atmospheric and solar neutrino problems
through vacuum oscillations,
\[ \mathcal{M}^L = M \left( \begin{array}{ccc}
0 & {1 \over \sqrt{2}} & {1 \over \sqrt{2}} \\
{1 \over \sqrt{2}} & {1 \over 2} & {-1 \over 2} \\
{1 \over \sqrt{2}} & {-1 \over 2} & {1 \over 2} \end{array} \right)
\begin{array}{c} \nu_e \\ \nu_\mu \\ \nu_\tau \end{array} \]
That is,
\[ ~~|\nu_e \rangle = \overbrace{{1 \over \sqrt{2}} |\nu_1 \rangle
- {1 \over \sqrt{2}} |\nu_2\rangle}^{\Delta \sim
\delta m^2_{solar}/2M}~(opp. CP) \]
\[ |\nu_\mu \rangle = {1 \over 2} |\nu_1 \rangle + {1 \over 2} |\nu_2 \rangle
+ {1 \over \sqrt{2}} |\nu_3 \rangle~~ \]
\[ |\nu_\tau \rangle = {1 \over 2} |\nu_1 \rangle + \underbrace{{1 \over 2} |\nu_2 \rangle 
- {1 \over \sqrt{2}} |\nu_3 \rangle}_{\Delta \sim \delta m^2_{atmos}/2M}~~ \]
where the two mass eigenstates comprising the $\nu_e$ interfere
in the $\beta \beta$ decay mass because they have opposite CP. \\
  
\noindent
This scenario has the following consequences: \\
$\bullet$ The $\beta \beta$ decay mass and solar neutrino mass
$\delta m^2$ are related by
$\langle \mathrm{m}^{Maj}_\nu \rangle = 
\delta m^2_{solar}$/4M, with the resulting rates for $\beta \beta$ 
decay thus being very small. \\
$\bullet$ The $\nu_\mu \rightarrow \nu_\tau$ oscillation is maximal
over terrestrial distances as $\nu_3$ beats against $\nu_1$ and
$\nu_2$; the small splitting of the latter two mass eigenstates
leads to 
$|\nu_e\rangle \rightarrow
(|\nu_\mu \rangle + |\nu_\tau \rangle)$ 
oscillations only over larger distances.  This solar neutrino oscillation
is also maximal. \\
$\bullet$ Interestingly, the MSW mechanism is not used. \\
This kind of mass matrix can arise naturally in model schemes,
as has been shown recently by Mohapatra and Nussinov \cite{mohapatra}.  Clearly it is just
one possibility among many, but suggests that the hints of
massive neutrinos we now have may yet conform to a simple 
pattern. \\

\section{Neutrinos and the r-process}
\noindent
As this talk is at an end, let me mention very quickly some
interesting connections between massive neutrinos and the 
explosive stellar synthesis of heavy nuclei by the rapid-neutron-capture or
r-process.  About half of the heavy elements above the iron
group are thought to be created by this process, where 
neutron capture is faster than $\beta$ decay, so that the
usual weak equilibrium condition of nuclei is replaced by
$(n,\gamma) \leftrightarrow (\gamma,n)$ equilibrium.  Thus the synthesis
occurs along a path through very neutron rich nuclei near
the neutron drip line. \\

\noindent
This process requires extraordinarily explosive conditions
\begin{center}
$\rho$(n) $\sim 10^{20}$ cm$^{-3}$~~~T $\sim 10^9$K~~~t $\sim$ 1sec
\end{center}
Probably the most plausible of the conjectured sites for the
r-process is the high-entropy, neutron rich gas near the mass
cut of a Type II supernova, the last material to be ejected.
As this material expands off the proto-neutron star, it 
undergoes an alpha-rich freezeout, and then an alpha process
that may continue to nuclei near A $\sim$ 100.  The result
is a soup of $\alpha$s, a few heavy seed nuclei, and excess
neutrons.  While detailed modeling of this ``hot bubble r-process"
fails in some details, the basic requirement of $\sim$ 100
neutrons per seed nucleus appears achievable \cite{woosley}.\\

\noindent
This matter experiences an enormous fluence of neutrinos,
emitted by the cooling protoneutron star.  As weak equilibrium
in maintained among the various neutrino species through
most of their random walk out of the neutron star, there is
an approximate equipartion of energy per flavor.  However,
the location of the neutrinosphere (the surface of last scattering)
does depend on flavor because of the strong, charged current
interactions of the $\nu_e$s and $\bar{\nu}_e$s.  Neutrinos
that decouple earlier do so at the higher ambient temperatures
characterizing the smaller neutrinospheres.  The results are
\begin{center}
T($\nu_\mu$,$\nu_\tau) \sim$ 8 MeV \\
T($\bar{\nu}_e) \sim$ 4.5 MeV \\
T($\nu_e) \sim$ 3.5 MeV
\end{center}
That is, the heavy flavor neutrinos are expected to be, on average,
significantly more energetic than the $\nu_e$s and $\bar{\nu}_e$s. \\
  
\begin{figure}[ht]
\psfig{bbllx=0.0cm,bblly=5cm,bburx=15.8cm,bbury=18.5cm,figure=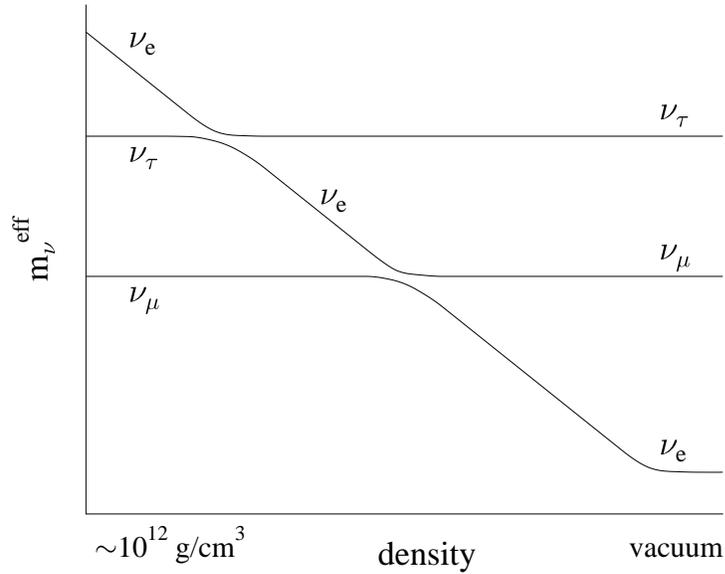,height=3in}
\caption{Three-flavor MSW diagram showing possible $\nu_e$--$\nu_\mu$
solar and $\nu_e$--$\nu_\tau$ supernova crossings.}
\label{three}
\end{figure}
  
\noindent
For the usual seesaw pattern of neutrino masses and a
cosmological interesting $\nu_\tau$ (i.e., a heavy neutrino 
with a mass in the neighborhood of 10 eV), the full MSW 
pattern is shown in Fig.~\ref{three}.  If the $\nu_e - \nu_\mu$
crossing is responsible for the solar neutrino problem,
a second crossing, $\nu_e - \nu_\tau$, is expected at a density
large compared to that of the solar core, but small compared
to the location of the supernova neutrinosphere 
($\sim 10^{12}$ g/cm$^3$).  For a very large range of mixing
angles, this crossing is adiabatic and thus leads to 
$\nu_e \leftrightarrow \nu_\tau$ conversion.  These spectra
thus change identities, leading to an anomalously hot $\nu_e$
flux from a Type II supernova. \\
  
\noindent
As the $\nu$-nucleon cross section is proportional to E$_\nu^2$,
\begin{center}
$\nu_e$ + n $\rightarrow$ e$^-$ + p~~~~is enhanced \\
$\bar{\nu}_e$ + p $\rightarrow$ e$^+$ + n~~is unchanged
\end{center}
For a rather extensive range of $\nu_e \leftrightarrow \nu_\tau$
mixing angles and $\delta m^2$, this crossing then destroys the
r-process: the hotter $\nu_e$s drive the matter proton rich \cite{qian}.
Thus, if one accepts this location as the site of the r-process,
very strong constraints on cosmologically interesting $\nu_\tau$s
are obtained. \\

\noindent
This provides a nice closing for this talk.  I began by discussing
how neutrino masses and other neutrino properties might tell us
about physics beyond the standard model - including, perhaps,
physics at the GUT scale.  This ending shows how
the pattern of neutrino masses may be equally relevant to our
low-energy world: the formation of large scale structure and
the existence of the transuranic elements may be issues connected
by the masses and mixing angles of neutrinos.\\

\noindent
This work was supported in part by the US Department of Energy.\\


\end{document}